\documentclass[reqno,10pt,dvips]{amsart}

\usepackage{amssymb,times,mathptmx,cite,eucal,psfrag,array,setspace,geometry}
\usepackage[dvips]{graphicx}

\numberwithin{equation}{section}

\newcolumntype{C}{>{$}c<{$}} 

\allowdisplaybreaks

\newcommand{\alg}[1]{\mathfrak{#1}}

\newcommand{\func}[2]{#1 \left( #2 \right)}

\newcommand{\brac}[1]{\left( #1 \right)}

\newcommand{\abs}[1]{\left| #1 \right|}

\newcommand{\ZZ}{\mathbb{Z}}

\newcommand{\ii}{\mathfrak{i}}

\newcommand{\bra}[1]{\bigl\langle #1 \bigr\rvert}
\newcommand{\ket}[1]{\bigl\lvert #1 \bigr\rangle}
\newcommand{\braket}[2]{\bigl\langle #1 \bigr\rvert \bigl. #2 \bigr\rangle}
\newcommand{\bracket}[3]{\bigl\langle #1 \bigr\rvert #2 \bigl\lvert #3 \bigr\rangle} 
\newcommand{\corrfn}[1]{\bigl\langle #1 \bigr\rangle}

\newcommand{\MinMod}[2]{\mathfrak{M} \left( #1 , #2 \right)}

\newcommand{\VerMod}[1]{\mathcal{V}_{#1}}
\newcommand{\IrrMod}[1]{\mathcal{L}_{#1}}
\newcommand{\IndMod}[1]{\mathcal{M}_{#1}}
\newcommand{\LogMod}[1]{\mathcal{I}_{#1}}

\newcommand{\fuse}{\times_{\! f}}

\newcommand{\eqnref}[1]{Equation~(\ref{#1})}
\newcommand{\eqnDref}[2]{Equations~(\ref{#1}) and (\ref{#2})}
\newcommand{\secref}[1]{Section~\ref{#1}}
\newcommand{\figref}[1]{Figure~\ref{#1}}
\newcommand{\tabref}[1]{Table~\ref{#1}}

\newcommand{\cft}{conformal field theory}
\newcommand{\cfts}{conformal field theories}
\newcommand{\ope}{operator product expansion}
\newcommand{\opes}{operator product expansions}
\newcommand{\hws}{highest weight state}

\newcommand{\hwm}{highest weight module}
\newcommand{\hwms}{highest weight modules}

\DeclareMathOperator{\id}{id}

\begin{document}

\title[From Percolation to $\log\:$CFT]{From Percolation to Logarithmic Conformal Field Theory}

\author[P Mathieu]{Pierre Mathieu}

\address[Pierre Mathieu]{
D\'{e}partement de Physique, de G\'{e}nie Physique et d'Optique \\
Universit\'{e} Laval \\
Qu\'{e}bec, Canada G1K 7P4
}

\email{pmathieu@phy.ulaval.ca}

\author[D Ridout]{David Ridout}

\address[David Ridout]{
D\'{e}partement de Physique, de G\'{e}nie Physique et d'Optique \\
Universit\'{e} Laval \\
Qu\'{e}bec, Canada G1K 7P4
}

\email{darid@phy.ulaval.ca}

\thanks{\today \\ This work is supported by NSERC}

\begin{abstract}
The smallest deformation of the minimal model $\MinMod{2}{3}$
that can accommodate Cardy's derivation of the percolation crossing
probability is presented.  It is shown that this leads to a consistent logarithmic \cft{} at $c=0$.  A simple recipe for computing the associated fusion rules is given.  The differences between this theory and the other recently proposed $c=0$ logarithmic \cfts{} are underlined.  The discussion also emphasises the existence of invariant logarithmic couplings that generalise Gurarie's anomaly number.
\end{abstract}

\maketitle

\onehalfspacing

\section{Introduction} \label{secIntro}

Percolation \cite{LanCon94,CarLec01} is one of the easiest of the statistical models to simulate numerically.  As such, it provides an excellent testing ground for uncovering how conformal invariance arises at critical points.  Upon varying the probability of a lattice site or bond to be open, one finds such a critical point delineating configurations in which one can or can not cross between opposite edges of the lattice via open sites or bonds.  At this critical point, percolation is believed to be described by a \cft{} with vanishing central charge, and this belief has been well tested through the determination of the quantities one can calculate within the theory and comparison with numerical simulations.  Naturally, the most important of these are the crossing probabilities, which give the probability that a random configuration will contain a cluster of open sites or bonds connecting opposite edges of the lattice.

Informed of the (then unpublished) numerical results of Langlands \emph{et al} \cite{LanUni92}, Aizenman suggested the conformal invariance of these crossing probabilities.  Upon being questioned on this, Cardy derived an exact closed-form expression for the horizontal crossing probability of a rectangular lattice in the thermodynamic limit (taken with the aspect ratio of the rectangle kept fixed), as a function of this aspect ratio.  The precise result \cite{CarCri92} is not relevant for the purposes of this paper, only that it is non-trivial (not constant).  However, we emphasise that the agreement with numerical simulation is impressive.  A rigorous proof of Cardy's result has since been presented \cite{SchSca00,SmiCri01}.

Cardy's derivation relied heavily on the machinery of \cft{}, hence may be viewed as a strong confirmation of the conformal invariance of critical percolation.  Paradoxically however, it has not been formulated within a completely coherent conformal-field-theoretic framework. 

Cardy interpreted the continuum limit of the percolation theory described above as a boundary \cft{} (on a rectangle) with vanishing central charge, and the horizontal crossing probability as (roughly speaking) a four-point correlation function on the upper half-plane $\corrfn{\func{\phi}{z_1} \func{\phi}{z_2} \func{\phi}{z_3} \func{\phi}{z_4}}$, involving a boundary field $\phi$ of conformal dimension $h=0$.  The role of the field $\func{\phi}{z_i}$ in the theory is to implement the change in the boundary conditions at $z_i$.  These two properties of $\phi$ (being boundary changing and having $h=0$) suggest its identification with the field $\phi_{1,2}$ in the minimal model $\MinMod{2}{3}$.  Then the null state $\brac{L_{-2} - \frac{3}{2} L_{-1}^2} \ket{\phi_{1,2}}$ determines the differential equation for the crossing probability in the usual manner \cite{BelInf84,DiFCon97}, and appropriate boundary conditions then select the required solution.

However, since $\MinMod{2}{3}$ is trivial, it is clear that it does not provide the proper framework in which to describe the above non-trivial four-point function.  Indeed, the field $\phi_{1,2}$ generates another null vector, $L_{-1} \ket{\phi_{1,2}}$, in the corresponding Verma module, which induces the differential equations $\partial_{z_i} \corrfn{\func{\phi_{1,2}}{z_1} \func{\phi_{1,2}}{z_2} \func{\phi_{1,2}}{z_3} \func{\phi_{1,2}}{z_4}} = 0$ (for all $i = 1, 2, 3, 4$).

It is not difficult to pinpoint the essential property that makes $\MinMod{2}{3}$ trivial and thereby implies the undesirable differential equations above.  To this end, let us examine a simple proof of this triviality.  As in any theory, the vacuum $\ket{0}$ must exist, and when $c=0$, the descendant $L_{-2} \ket{0}$ is null\footnote{We will freely use the terms \emph{null vector} (a state of zero-norm), \emph{singular vector} (a descendant state annihilated by $L_1$ and $L_2$) and \emph{principal singular vector} (a singular vector which is not itself a descendant of a singular vector).  Such states may or may not identically vanish, and we shall refer to non-vanishing states as \emph{physical}.}.  The corresponding null field is then the energy-momentum tensor $\func{T}{z}$ whose modes $L_n$ must therefore annihilate all physical states.  This implies that the only physical state is the vacuum itself.

Let us reformulate this result in a more mathematically precise manner:  When $c=0$, the only physical state which can coexist with the \emph{irreducible} vacuum module is the vacuum itself\footnote{We recall that a module is \emph{reducible} if it contains a non-trivial submodule and \emph{decomposable} if it can be written as the direct sum of two non-trivial submodules.  \emph{Irreducible} and \emph{indecomposable} describe the opposite situations, respectively.}.  This irreducibility condition forces the modes of $\func{T}{z}$ to act as the zero operator on the physical state space, and it is therefore this very condition that Cardy's result forces us to relax.  We will explore the consequences of breaking the hypothesis of an irreducible vacuum module in the following sections, and show that this simple act leads to a consistent \cft{} in which the non-triviality of the $\phi_{1,2}$ four-point function is fact.  This theory is constructed from the minimal set of conditions ensuring this non-triviality, and will turn out to be a \emph{logarithmic} \cft{}.

\section{Heuristic Considerations} \label{secConstructions1}

It proves convenient to fix a few notations from the outset.  We present a part of the extended Kac table for $c=0$ in \tabref{tabExtKacc=0}, in which the dimensions $h_{r,s}$ of the (possibly primary) fields $\phi_{r,s}$ are displayed for $r=1,2,3$ and $s=1,\ldots,10$.  This extends the Kac table of the minimal model $\MinMod{2}{3}$.  We will denote the Verma module generated from the state $\ket{\phi_{r,s}}$ by $\VerMod{r,s}$ and its irreducible quotient by $\IrrMod{r,s}$.  Note that the $\VerMod{r,s}$ with $r$ even or $s$ a multiple of $3$ have their maximal submodules generated by a single singular vector at grade $rs$, whereas the maximal submodules of the other $\VerMod{r,s}$ associated to the extended Kac table are generated by two singular vectors at grades $rs$ and $\brac{r-2} \brac{s-3}$, respectively \cite{FeiVer84}.  We will also be interested in the indecomposable (but reducible) modules given by quotienting each $\VerMod{r,s}$ by the Verma module generated by the singular vector at grade $rs$.  These modules will be denoted by $\IndMod{r,s}$.

\begin{table}
\begin{center}
\setlength{\extrarowheight}{4pt}
\begin{tabular}{|C|C|C|C|C|C|C|C|C|C|C}
\hline
0 & 0 & \tfrac{1}{3} & 1 & 2 & \tfrac{10}{3} & 5 & 7 & \tfrac{28}{3} & 12 & \cdots \\[1mm]
\hline
\tfrac{5}{8} & \tfrac{1}{8} & \tfrac{-1}{24} & \tfrac{1}{8} & \tfrac{5}{8} & \tfrac{35}{24} & \tfrac{21}{8} & \tfrac{33}{8} & \tfrac{143}{24} & \tfrac{65}{8} & \cdots \\[1mm]
\hline
2 & 1 & \tfrac{1}{3} & 0 & 0 & \tfrac{1}{3} & 1 & 2 & \tfrac{10}{3} & 5 & \cdots \\[1mm]
\hline
\end{tabular}
\vspace{3mm}
\caption{The first three rows of the extended Kac table for $c=0$, displaying the conformal dimensions $h_{r,s} = \bigl( \brac{3r-2s}^2 - 1 \bigr) / 24$ of the fields $\phi_{r,s}$.  $r$ increases downwards, and $s$ increases to the right, and the top-left-hand corner corresponds to the identity field $\phi_{1,1}$, which with $\phi_{1,2}$ exhausts the usual Kac table for $\MinMod{2}{3}$.  In fact, all dimensions appearing in the extended Kac table may be found in the first two semi-infinite rows, as $h_{r,s} = h_{r+2,s+3} = h_{r,3r-s}$.} \label{tabExtKacc=0}
\end{center}
\end{table}

We begin with the observation that translation invariance of the vacuum requires that $L_{-1} \ket{0} = 0$, and this of course is reinforced by the state-field correspondence:  $\ket{0} \leftrightarrow I$ so that $L_{-1} \ket{0} \leftrightarrow \partial I = 0$.  Since we have already argued that the vacuum module cannot be irreducible, the only remaining possibility is that the vacuum module is the indecomposable (but not irreducible) $\IndMod{1,1} = \VerMod{1,1} / \VerMod{1,4}$.  In other words, we require that the singular vector $L_{-2} \ket{0}$ be \emph{non-vanishing}, and in this way recover a non-trivial (though null) energy-momentum tensor $\func{T}{z}$.

Furthermore, Cardy's result relies upon the identification of his $h=0$ boundary field with $\phi_{1,2}$.  Indeed, we want to be able to derive the differential equation induced by the descendant singular vector at grade $2$, but not be able to derive the differential equations induced by the singular vector at grade $1$.  We propose to achieve this by forcing the singular vector at grade $1$ to be non-vanishing, and its grade $2$ counterpart to vanish identically.  Accommodating Cardy's result then requires also taking the physical module corresponding to the primary field $\phi_{1,2}$ to be indecomposable (but not irreducible):  $\IndMod{1,2} = \VerMod{1,2} / \VerMod{1,5}$.

\begin{figure}
\begin{center}
\includegraphics[width=7cm]{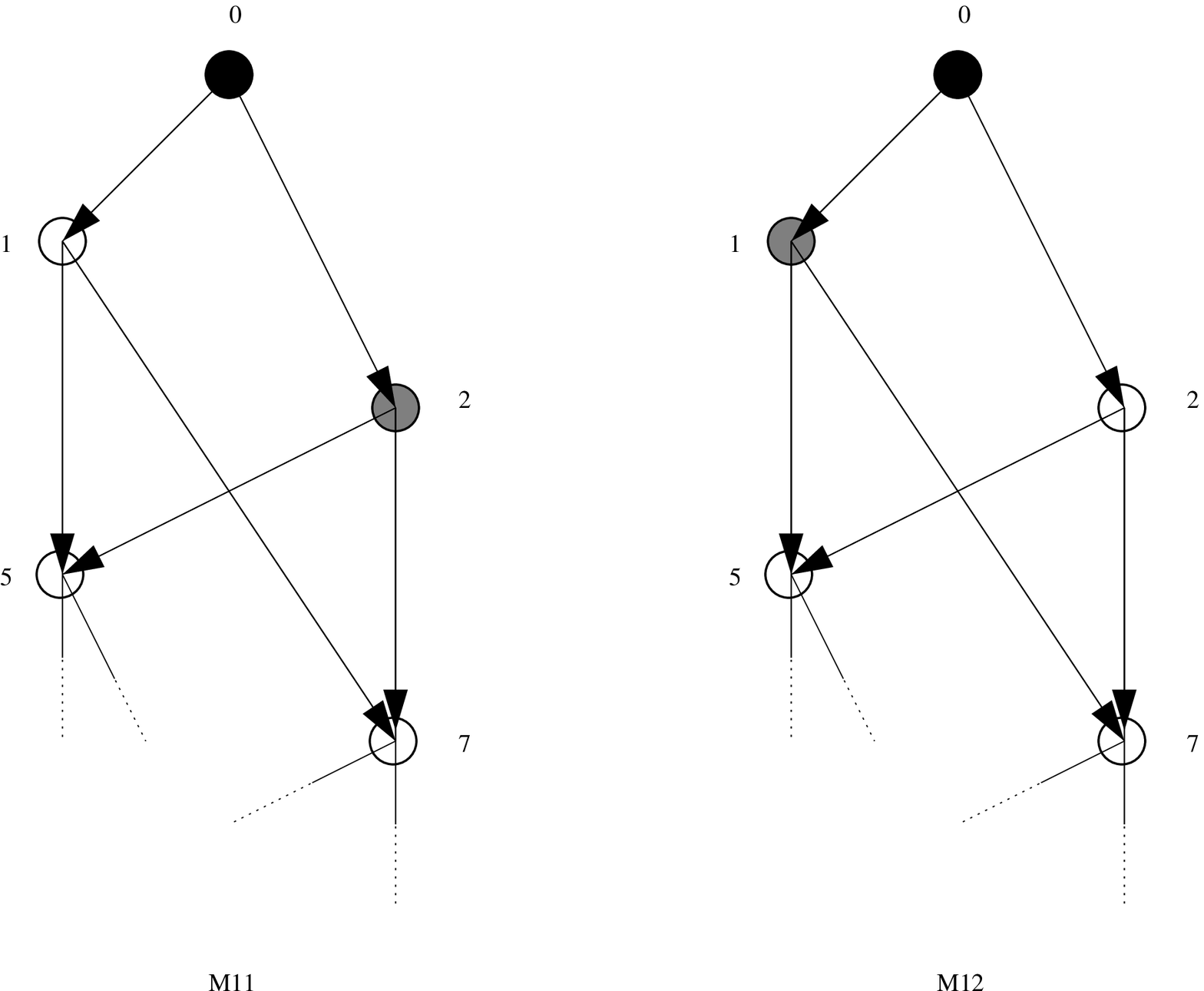}
\caption{A schematic picture of the physical modules of conformal dimension $0$ in our $c=0$ theory.  The black circles represent the highest weight states, grey denotes a singular vector that does \emph{not} identically vanish, and white denotes the identically vanishing singular vectors.  These states are labelled by their conformal dimension.} \label{figModulesh=0}
\end{center}
\end{figure}

We therefore see that in order to put Cardy's derivation in a consistent conformal-field-theoretic framework, we must start with two reducible but indecomposable modules of highest weight $h=0$.  These are illustrated schematically in \figref{figModulesh=0}.  The corresponding primary fields are distinguished by their different descendant structures, and in this way the Kac symmetry of $\MinMod{2}{3}$ is broken:  $\phi_{1,1} \neq \phi_{1,2}$.

We emphasise that what we have described amounts to a \emph{minimal fit} in that all of our reasoning has been forced by one goal---validating Cardy's derivation, itself validated conclusively by numerical simulations.  It remains to ``flesh out'' this theory and check its consistency, thus verifying that the formalism we construct achieves our goal.

\psfrag{0}[][]{$0$}
\psfrag{1}[][]{$1$}
\psfrag{2}[][]{$2$}
\psfrag{5}[][]{$5$}
\psfrag{7}[][]{$7$}
\psfrag{M11}[][]{$\IndMod{1,1}$}
\psfrag{M12}[][]{$\IndMod{1,2}$}

In the remainder of this section, we will explore the theory we are constructing in a somewhat heuristic manner, so as to quickly deduce certain necessary features.  In the following section, we will revisit our constructions using more precise analysis techniques, and thereby prove that these necessary features are indeed present.  It is these precise methods which will uncover the logarithmic structure of the percolation \cft{}.

For now, we explore the field content of the theory generated by the modules $\IndMod{1,1}$ and $\IndMod{1,2}$.  The vanishing singular vector $\brac{L_{-2} - \frac{3}{2} L_{-1}^2} \ket{\phi_{1,2}} = 0$ implies, via the usual consideration of three-point functions \cite{BelInf84,DiFCon97}, the fusion rules
\begin{equation} \label{eqnFR12byrs}
\phi_{1,2} \times \phi_{r,s} = \phi_{r,s-1} + \phi_{r,s+1},
\end{equation}
where $\phi_{r,0}$ is formally set to zero.  When the module generated by $\phi_{1,2}$ is irreducible, the other vanishing singular vector further constrains the fields appearing in the above fusion rule.  We will proceed however, by \emph{assuming} that in the indecomposable case, this other singular vector $L_{-1} \ket{\phi_{1,2}}$ (which is non-vanishing) does \emph{not} lead to constraints on the above fusion rules.  This assumption will keep our conclusions in this section on a heuristic level, but it will be validated in the more precise treatment of the following section.

So, accepting the fusion rules (\ref{eqnFR12byrs}) for the moment, we obviously have $\phi_{1,2} \times \phi_{1,2} = \phi_{1,1} + \phi_{1,3}$.  The field $\phi_{1,3}$ must appear on the right hand side if the $\phi_{1,2}$ four-point function is to be non-trivial.  This follows from
\begin{equation} \label{eqnCF12.12.12.d12}
\corrfn{\func{\phi_{1,2}}{z_1} \func{\phi_{1,2}}{z_2} \func{\phi_{1,2}}{z_3} \func{\brac{L_{-1} \phi_{1,2}}}{z_4}} = \partial_{z_4} \corrfn{\func{\phi_{1,2}}{z_1} \func{\phi_{1,2}}{z_2} \func{\phi_{1,2}}{z_3} \func{\phi_{1,2}}{z_4}} \neq 0.
\end{equation}
If $\phi_{1,3}$ did not appear on the right hand side of this fusion rule, then inserting the corresponding \ope{} as $z_1 \rightarrow z_2$ and then again as $z_2 \rightarrow z_3$ would reduce the correlation functions in \eqnref{eqnCF12.12.12.d12} to a linear combination of two-point functions, each of which involves a descendant of $\func{\phi_{1,2}}{z_3}$ and $\func{\brac{L_{-1} \phi_{1,2}}}{z_4} = \func{\partial \phi_{1,2}}{z_4}$.  But these two-point functions all vanish, as
$h_{1,2} = 0$ implies that $\corrfn{\func{\phi_{1,2}}{z_3} \func{\phi_{1,2}}{z_4}} = 1$.  We conclude then that the presence of $\phi_{1,3}$ in the theory is necessary.

Consider now the fusion rule $\phi_{1,2} \times \phi_{1,3} = \phi_{1,2} + \phi_{1,4}$.  Inserting the \opes{} corresponding to (\ref{eqnFR12byrs}) with $r=1,s=2$ (as $z_1 \rightarrow z_2$) and then again with $r=1,s=3$ (as $z_2 \rightarrow z_3$) into \eqnref{eqnCF12.12.12.d12}, we obtain a linear combination of two-point functions involving the null field $\func{\partial \phi_{1,2}}{z_4}$ and descendants of $\func{\phi_{1,2}}{z_3}$ or $\func{\phi_{1,4}}{z_3}$.  As we know, those involving $\phi_{1,2}$-descendants vanish, so global conformal invariance and $h_{1,4} = 1$ (see \tabref{tabExtKacc=0}) imply that the non-vanishing contributions are obtained from the action of differential operators (with respect to $z_3$) acting on
\begin{equation} \label{eqnCF14.d12}
\corrfn{\func{\phi_{1,4}}{z_3} \func{\partial \phi_{1,2}}{z_4}} = \frac{C}{\brac{z_3 - z_4}^2} \qquad \text{(for some constant $C$)}.
\end{equation}
We remark that $\func{\partial \phi_{1,2}}{z_4}$ is a primary (though null) field of dimension $1$, and non-triviality requires $C \neq 0$.  This in turn obviously requires that $\phi_{1,4}$ belong to the theory.

Summarising, the difference between the trivial theory constructed from irreducible modules and the theory we are constructing here from indecomposable (but reducible) modules is (at this level of rigour) that in the latter case, the presence of $\phi_{1,3}$ opens up a new channel in the \opes{}, which allows the possibility of a non-trivial four-point function $\corrfn{\phi_{1,2} \phi_{1,2} \phi_{1,2} \phi_{1,2}}$.

We point out that $\corrfn{\func{\partial \phi_{1,2}}{z_3} \func{\partial \phi_{1,2}}{z_4}} = \partial_{z_3} \partial_{z_4} \corrfn{\func{\phi_{1,2}}{z_3} \func{\phi_{1,2}}{z_4}} = 0$, hence that $\phi_{1,4} \neq \partial \phi_{1,2}$, by \eqnref{eqnCF14.d12}.  Note that \eqnref{eqnCF14.d12} does imply that $\ket{\phi_{1,4}}$ and $L_{-1} \ket{\phi_{1,2}}$ have non-zero inner-product, indicating that these states both belong to some common indecomposable module (this refines an observation of \cite{SimPer07}).  We will see shortly that this is the case, and it is the logarithmic structure of this module which makes it possible.

We could continue this process, generating $\phi_{1,5}$ and beyond, but as we have already mentioned, this all relies on the assumption that the fusion rules (\ref{eqnFR12byrs}) are correct.  Justifying this assumption is somewhat delicate because we are working with modules more general than the familiar irreducible ones, so the usual methods of inferring fusion rules (examining the action of null vectors on three-point functions in particular) might not be valid.  The key point here is that we want a method in which we can distinguish between vanishing and non-vanishing null-vectors, and we expect that this will not be easy if inner-products and correlation functions are used.  We therefore turn to a direct \emph{algebraic} computation of these fusion rules which make no reference to correlation functions and inner-products.

\section{Fusion Rules and the Rise of Logarithms} \label{secConstructions2}

To investigate the fusion ring generated by the indecomposable modules $\IndMod{1,1}$ and $\IndMod{1,2}$ from which the percolation \cft{} is constructed, we turn to the algorithm of Nahm.  This was originally introduced in \cite{NahQua94} for so-called quasi-rational modules, and was extended (and made more transparent) by Gaberdiel and Kausch \cite{GabInd96} using earlier results of Gaberdiel \cite{GabFus94}.  We shall not discuss the details of this algorithm here.  We only mention that it provides information on the decomposition of the fusion of two modules, by utilising a natural representation of the chiral symmetry algebra on the set of \opes{} (for primary \emph{and} descendant fields) corresponding to the states in these modules.  Importantly, the vanishing singular vectors of the modules to be fused are inputs to this algorithm, and at no point do we use the inner-products on the modules.

Of course, there are infinitely many such \opes{}, as there are an infinite number of descendant states in each module, graded by their conformal dimensions.  It is possible however to consistently truncate this set of \opes{} to a finite number, by imposing an upper-bound on the grade, relative to the \hws{} (mathematically, one considers the elements of an appropriate filtration of the module).  Of course, this means that one only obtains a finite amount of information concerning the structure of the decomposition of the fused modules.  Fortuitously, one can deduce the entire decomposition structure from such a (sufficiently large) finite amount of information, essentially by ``looking deeply enough'' to see the principal singular vectors (or not, as the case may be).  It is this feature that makes the Nahm-Gaberdiel-Kausch algorithm (whose practical implementation is nicely detailed in \cite{GabInd96}) so powerful (and general).

We illustrate the application of this algorithm to the fusion of the indecomposable module $\IndMod{1,2}$ with itself (as expected, the indecomposable vacuum module $\IndMod{1,1}$ still acts as the identity of the fusion ring).  A theorem of Nahm \cite{NahQua94} guarantees that the zero-grade states in the decomposition of the fused modules can be associated with the states in a two-dimensional Cartesian product space\footnote{Generally, they would be associated with a subspace of these states and one would have to search for the \emph{spurious subspace} \cite{NahQua94}.  However, there are no spurious states in this case.}.  Computing the natural representative for $L_0$ (see \cite{GabFus94} for explicit formulae) on this space gives a matrix form for this representative:
\begin{equation}
L_0 = 
\begin{pmatrix}
0 & 0 \\
1 & \tfrac{1}{3}
\end{pmatrix}
\quad \text{with respect to the ordered basis} \quad
\begin{Bmatrix}
\ket{\phi_{1,2}} \times \ket{\phi_{1,2}} \\
L_{-1} \ket{\phi_{1,2}} \times \ket{\phi_{1,2}}
\end{Bmatrix}
.
\end{equation}
Thus $L_0$ is diagonalisable with eigenvalues $0$ and $1/3$ on the zero-grade states of the fusion of $\IndMod{1,2}$ with itself.  This is perfectly consistent with the fusion rule (\ref{eqnFR12byrs}) with $r=1,s=2$.

To completely identify the character of the modules appearing in the decomposition of this fusion process, we must repeat this computation whilst considering all states up to grade $3$.  This time, we compute\footnote{There is in addition a one-dimensional spurious subspace to be determined in this case.  We used the method suggested in \cite{GabInd96} to find it.} a $9 \times 9$ representing matrix for $L_0$, which turns out to be diagonalisable with eigenvalues $0$, $2$, $3$, $\tfrac{1}{3}$, $\tfrac{4}{3}$, $\tfrac{7}{3}$, $\tfrac{7}{3}$, $\tfrac{10}{3}$, and $\tfrac{10}{3}$.  This result is only consistent with the fusion decomposition
\begin{equation}
\IndMod{1,2} \fuse \IndMod{1,2} = \IndMod{1,1} \oplus \IndMod{1,3},
\end{equation}
where we denote the fusion operation by $\fuse$, to distinguish it from the Cartesian product ($\oplus$ denotes the direct sum of modules, as always).  This is the precise version of the fusion rule (\ref{eqnFR12byrs}) (with $r=1,s=2$) which we proposed heuristically in \secref{secConstructions1}.  We mention that $\IndMod{1,3} = \IrrMod{1,3}$ is in fact irreducible.

A more interesting computation is to determine the fusion of $\IndMod{1,2}$ and $\IndMod{1,3}$ to grade $1$.  By Gaberdiel and Kausch's generalisation of Nahm's theorem to all grades \cite{GabInd96}, we compute within a four dimensional space, finding\footnote{Again, there are no spurious states in this case.}
\begin{equation}
L_0 = 
\begin{pmatrix}
\tfrac{1}{3} & 0 & \tfrac{2}{9} & \tfrac{8}{27} \\
0 & \tfrac{4}{3} & \tfrac{2}{3} & \tfrac{4}{9} \\
1 & 0 & \tfrac{4}{3} & 0 \\
0 & 1 & 0 & 1
\end{pmatrix}
\quad \text{with respect to the ordered basis} \quad
\begin{Bmatrix}
\ket{\phi_{1,2}} \times \ket{\phi_{1,3}} \\
L_{-1} \ket{\phi_{1,2}} \times \ket{\phi_{1,3}} \\
\ket{\phi_{1,2}} \times L_{-1} \ket{\phi_{1,3}} \\
L_{-1} \ket{\phi_{1,2}} \times L_{-1} \ket{\phi_{1,3}}
\end{Bmatrix},
\end{equation}
which turns out \emph{not} to be diagonalisable.  Indeed, it has simple eigenvalues $0$ and $2$ and a Jordan cell of rank $2$ corresponding to the eigenvalue $1$.  Computing the action of $L_{-1}$ in the same way, we find that the eigenstate of eigenvalue $0$ is mapped to the true eigenstate of eigenvalue $1$ by $L_{-1}$ whereas its Jordan partner is mapped to the eigenstate of eigenvalue $2$.  This suggests the identification of the eigenstates of eigenvalues $0$, $1$ and $2$ with $\ket{\phi_{1,2}}$, $L_{-1} \ket{\phi_{1,2}}$ and $L_{-1} \ket{\phi_{1,4}}$, respectively, where $\ket{\phi_{1,4}}$ denotes the Jordan partner to $L_{-1} \ket{\phi_{1,2}}$.  We normalise this partner state so that
\begin{equation} \label{eqnNormJord12}
L_0 \ket{\phi_{1,4}} = \ket{\phi_{1,4}} + L_{-1} \ket{\phi_{1,2}},
\end{equation}
fixing it up to multiples of $L_{-1} \ket{\phi_{1,2}}$.

\begin{figure}
\begin{center}
\includegraphics[width=14cm]{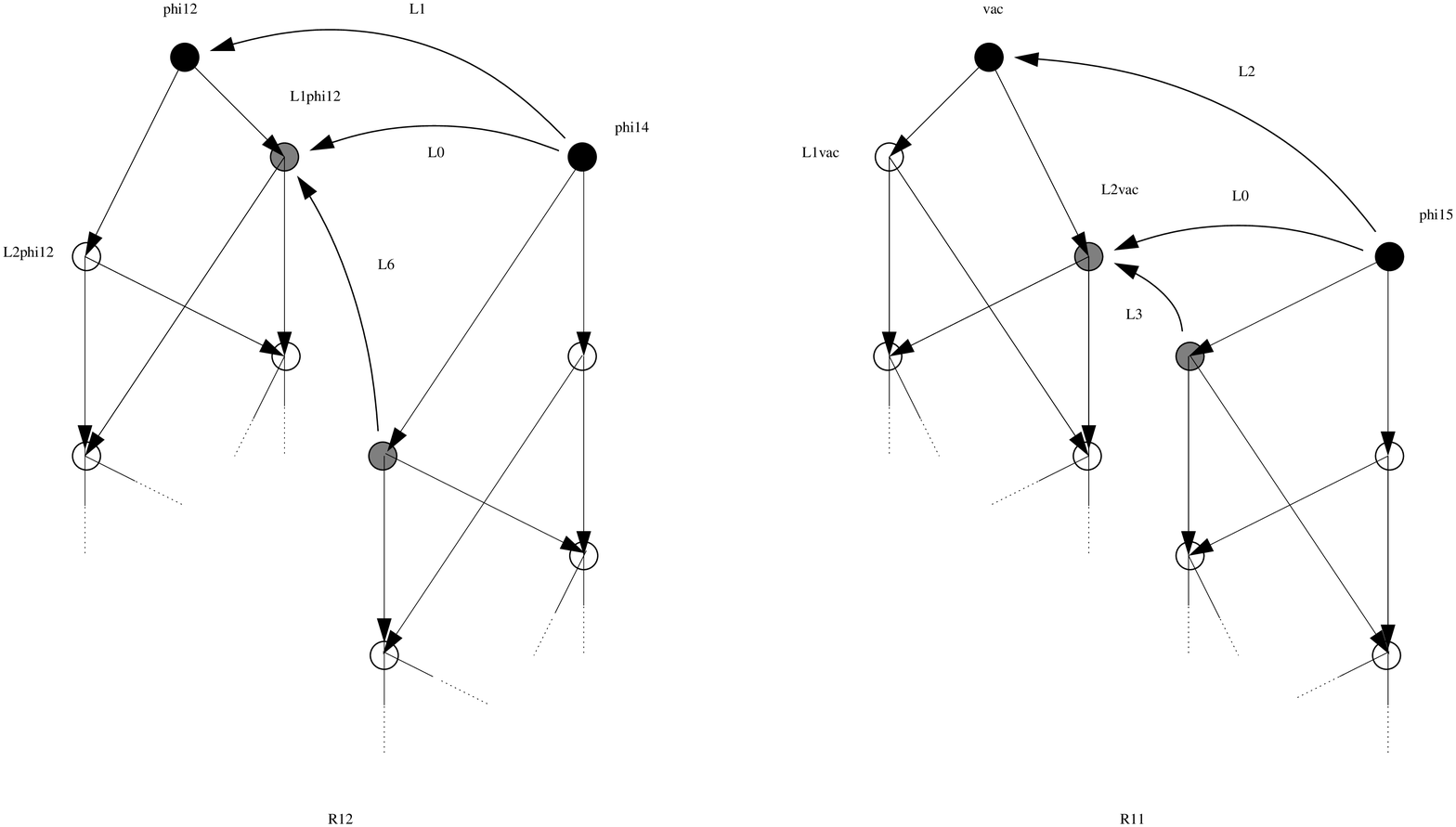}
\caption{A schematic picture of the staggered modules $\LogMod{1,4}$ and $\LogMod{1,5}$ showing the singular vector structure of the two \hwms{} from which they are constructed.  White circles correspond to identically vanishing singular vectors, whereas grey indicate that the singular vector is non-vanishing.  Here, $\ket{\chi}$ is the vanishing singular vector $\brac{L_{-2} - \frac{3}{2} L_{-1}^2} \ket{\phi_{1,2}}$, and $A_3$ is defined after \eqnref{eqnA3.17}.  The curved arrows depict (roughly) how the Virasoro mode action ``glues'' these modules together to form the staggered module (the precise actions are given in the text).} \label{figStagMods}
\end{center}
\end{figure}

Let $\LogMod{1,4}$ denote the module obtained from fusing $\IndMod{1,2}$ and $\IndMod{1,3}$:
\begin{equation}
\IndMod{1,2} \fuse \IndMod{1,3} = \LogMod{1,4}.
\end{equation}
A full picture of the structure of this module requires computing to grade $6$, so as to ``see'' all principal singular vectors.  This is computationally intensive, but straight-forward to program (we used \textsc{Maple}).  The result is that $\LogMod{1,4}$ is the \emph{vector space} direct sum of the modules $\IndMod{1,2}$ and $\IndMod{1,4}$, but is \emph{indecomposable} itself as a Virasoro module\footnote{The notation $\LogMod{}$ emphases the indecomposable aspect of these modules and is used instead of the more familiar $\mathcal{R}$  which stresses their reducibility.}.  This is an example of a \emph{staggered module}, in the terminology of Rohsiepe \cite{RohRed96}:  $\LogMod{1,4}$ has a submodule isomorphic to the \hwm{} $\IndMod{1,2}$ and its quotient by this submodule is isomorphic to the \hwm{} $\IndMod{1,4}$.  Mathematically, this is summarised by the exact sequence
\begin{equation}
0 \longrightarrow \IndMod{1,2} \longrightarrow \LogMod{1,4} \longrightarrow \IndMod{1,4} \longrightarrow 0.
\end{equation}
We illustrate $\LogMod{1,4}$ schematically in \figref{figStagMods}.  Note, however, that $\LogMod{1,4}$ is not itself a \hwm{}.

\psfrag{vac}[][]{$\scriptstyle \left| 0 \right\rangle$}
\psfrag{L1vac}[][]{$\scriptstyle L_{-1} \left| 0 \right\rangle$}
\psfrag{L2vac}[][]{$\scriptstyle L_{-2} \left| 0 \right\rangle$}
\psfrag{L0}[][]{$\scriptstyle L_0$}
\psfrag{L1}[][]{$\scriptstyle L_1$}
\psfrag{L2}[][]{$\scriptstyle L_2$}
\psfrag{L3}[][]{$\scriptstyle A_3$}
\psfrag{L6}[][]{$\scriptstyle L_6 + \ldots$}
\psfrag{phi12}[][]{$\scriptstyle \left| \phi_{1,2} \right\rangle$}
\psfrag{phi14}[][]{$\scriptstyle \left| \phi_{1,4} \right\rangle$}
\psfrag{phi15}[][]{$\scriptstyle \left| \phi_{1,5} \right\rangle$}
\psfrag{L1phi12}[][]{$\scriptstyle L_{-1} \left| \phi_{1,2} \right\rangle$}
\psfrag{L2phi12}[][]{$\scriptstyle \left| \chi \right\rangle$}
\psfrag{R11}[][]{$\LogMod{1,5}$}
\psfrag{R12}[][]{$\LogMod{1,4}$}

In fact, $\LogMod{1,4}$ is generated by the state $\ket{\phi_{1,4}}$, as computing $L_1$ with the Nahm-Gaberdiel-Kausch algorithm gives
\begin{equation} \label{eqnL1.14}
L_1 \ket{\phi_{1,4}} = \frac{-1}{2} \ket{\phi_{1,2}}.
\end{equation}
This non-trivial relation does not follow from \eqnref{eqnNormJord12} and the Virasoro commutation relations, and in fact serves to fix the structure of the staggered module $\LogMod{1,4}$ completely.  Note that upon quotienting by $\IndMod{1,2}$, we recover\footnote{In particular, note that the vanishing grade $4$ singular vector $\ket{\zeta}$ descended from $\ket{\phi_{1,4}}$ is not the $\IndMod{1,4}$ singular vector
\begin{equation*}
\brac{L_{-4} - L_{-3} L_{-1} - L_{-2}^2 + \frac{5}{3} L_{-2} L_{-1}^2 - \frac{1}{4} L_{-1}^4} \ket{\phi_{1,4}},
\end{equation*}
as one might have expected.  Solving $L_1 \ket{\zeta} = L_2 \ket{\zeta} = 0$ in $\LogMod{1,4}$ (subject to the vanishing of the $\IndMod{1,2} \subset \LogMod{1,4}$ singular vector $\brac{L_{-2} - \frac{3}{2} L_{-1}^2} \ket{\phi_{1,2}}$), yields the true (identically vanishing) $\LogMod{1,4}$ singular vector
\begin{equation*}
\ket{\zeta} = \brac{L_{-4} - L_{-3} L_{-1} - L_{-2}^2 + \frac{5}{3} L_{-2} L_{-1}^2 - \frac{1}{4} L_{-1}^4} \ket{\phi_{1,4}} + \brac{\frac{1}{2} L_{-5} + \frac{4}{3} L_{-4} L_{-1} - \frac{8}{9} L_{-3} L_{-2}} \ket{\phi_{1,2}} = 0.
\end{equation*}
Of course this reduces to the $\IndMod{1,4}$ singular vector upon quotienting by $\IndMod{1,2}$.} the highest weight condition for $\ket{\phi_{1,4}} \in \IndMod{1,4}$.

We are now in a position to verify 
\eqnref{eqnCF14.d12}, which we showed in \secref{secConstructions1} was necessary for the non-triviality of the $\phi_{1,2}$ four-point function.  It is now clear that the constant $C$ appearing there is just
\begin{equation} \label{eqnLogCoup14}
C = \bracket{\phi_{1,4}}{L_{-1}}{\phi_{1,2}} = \frac{-1}{2} \braket{\phi_{1,2}}{\phi_{1,2}} = \frac{-1}{2},
\end{equation}
using \eqnref{eqnL1.14}.  We remark that even though $L_{-1} \ket{\phi_{1,2}}$ is null, it can still have a non-vanishing inner-product with another state (in particular its Jordan partner state $\ket{\phi_{1,4}}$).  This would not be possible in a \hwm{}, so we see in hindsight that Cardy's derivation can only be valid in a \cft{} based on modules more general than \hwms{} (such as the staggered module $\LogMod{1,4}$ we have discovered here).  In other words, this exposes clearly the necessity of having a non-diagonalisable $L_0$ in the percolation \cft{}.

As is well known, non-diagonalisability of $L_0$ is often taken as a defining property of logarithmic \cfts{} \cite{GurLog93}.  This logarithmic structure is easy to elucidate in the present case.  First, \eqnDref{eqnNormJord12}{eqnL1.14} allow us to derive the \ope{}
\begin{equation}
\func{T}{z} \func{\phi_{1,4}}{w} = \frac{-1}{2} \frac{\func{\phi_{1,2}}{w}}{\brac{z-w}^3} + \frac{\func{\phi_{1,4}}{w} + \func{\partial \phi_{1,2}}{w}}{\brac{z-w}^2} + \frac{\func{\partial \phi_{1,4}}{w}}{z-w} + \ldots
\end{equation}
This and global conformal invariance of the vacuum now imply that $\corrfn{\func{\phi_{1,4}}{z} \func{\phi_{1,4}}{w}}$ satisfies the differential equation
\begin{equation} \label{eqnCF14.14}
\brac{z \partial_z + w \partial_w + 2} \corrfn{\func{\phi_{1,4}}{z} \func{\phi_{1,4}}{w}} = \frac{1}{\brac{z-w}^2} \qquad \Rightarrow \qquad \corrfn{\func{\phi_{1,4}}{z} \func{\phi_{1,4}}{w}} = \frac{A + \log \brac{z-w}}{\brac{z-w}^2}
\end{equation}
where $A$ is some constant\footnote{Note that a direct consequence of the logarithm appearing in \eqnref{eqnCF14.14} is that the (standard) inner-product $\braket{\phi_{1,4}}{\phi_{1,4}}$ diverges.  Indeed, considering $L_{-1} \ket{\phi_{1,2}}$ and its Jordan partner $\ket{\phi_{1,4}}$, the norm of the former vanishes and that of the latter diverges, but their inner-product is finite and non-zero (\eqnref{eqnLogCoup14}).  This reflects the simple fact that there is no single invariant inner-product defined on these non-\hwms{}.  Note that if the norm of $\ket{\phi_{1,4}}$ were not divergent, then letting $L_0$ act on the bra and ket respectively in $\bracket{\phi_{1,4}}{L_0}{\phi_{1,4}}$ would lead to
\begin{equation*}
\braket{\phi_{1,4}}{\phi_{1,4}} = \braket{\phi_{1,4}}{\phi_{1,4}} + \bracket{\phi_{1,4}}{L_{-1}}{\phi_{1,2}} = \braket{\phi_{1,4}}{\phi_{1,4}} - \frac{1}{2},
\end{equation*}
a contradiction.  Here, it is important to note that $L_0^{\dag} \ket{\phi_{1,4}} = \ket{\phi_{1,4}}$ (and $L_0^{\dag} L_{-1} \ket{\phi_{1,2}} = L_{-1} \ket{\phi_{1,2}} + \ket{\phi_{1,4}}$).  Thus, $L_0 \neq L_0^{\dag}$ as required by non-diagonalisability.}.  In fact, we can set $A = 0$ because we still have the freedom to redefine $\phi_{1,4}$ as $\phi_{1,4} + a \partial \phi_{1,2}$ for arbitrary $a$, without affecting the defining \eqnDref{eqnNormJord12}{eqnL1.14}.

This discussion firmly establishes the theory we are constructing as the conformal field theory associated to critical percolation by Cardy.  However, we have not yet exhausted the richness of this theory.  In particular, we can apply the algorithm of Nahm, Gaberdiel and Kausch to the fusion of $\IndMod{1,3} = \IrrMod{1,3}$ with itself.  Despite this module being irreducible, we still compute non-diagonalisable representatives for $L_0$ on the fusion product, and by computing to grade $5$, we conclude that
\begin{equation}
\IndMod{1,3} \fuse \IndMod{1,3} = \IndMod{1,3} \oplus \LogMod{1,5}.
\end{equation}
Here, $\LogMod{1,5}$ is another staggered module, structurally described by the exact sequence $0 \rightarrow \IndMod{1,1} \rightarrow \LogMod{1,5} \rightarrow \IndMod{1,5} \rightarrow 0$.  The field $\phi_{1,5}$ is the Jordan partner of the energy-momentum tensor $T$:
\begin{equation} \label{eqnL0.15}
L_0 \ket{\phi_{1,5}} = 2 \ket{\phi_{1,5}} + L_{-2} \ket{0},
\end{equation}
and computing the action of $L_2$ on $\ket{\phi_{1,5}}$ gives
\begin{equation} \label{eqnL2.15}
L_2 \ket{\phi_{1,5}} = \frac{-5}{8} \ket{0}.
\end{equation}
We illustrate this module schematically in \figref{figStagMods}.

Again, this staggered module structure leads to the appearance of logarithms in correlation functions.  \eqnDref{eqnL0.15}{eqnL2.15} imply the \ope{}
\begin{equation}
\func{T}{z} \func{\phi_{1,5}}{w} = \frac{-5}{8} \frac{1}{\brac{z-w}^4} + \frac{2 \func{\phi_{1,5}}{w} + \func{T}{w}}{\brac{z-w}^2} + \frac{\func{\partial \phi_{1,5}}{w}}{z-w} + \ldots,
\end{equation}
and the differential equations derived from the conformal invariance of the vacuum yield
\begin{equation} \label{eqnCF15.15}
\corrfn{\func{\phi_{1,5}}{z} \func{\phi_{1,5}}{w}} = \frac{5}{4} \frac{\log \brac{z-w}}{\brac{z-w}^4}.
\end{equation}
Here, we have redefined $\func{\phi_{1,5}}{z}$ so as to set the arbitrary constant coming from the differential equation to zero (as discussed after \eqnref{eqnCF14.14}).  We see immediately that the norm of $\ket{\phi_{1,5}}$ also diverges.

Thus far, we have constructed a part of the spectrum of a \cft{} consistent with Cardy's percolation result.  Of course, it is possible to continue the analysis, uncovering more of this percolation \cft{} structure.  We have computed several more fusion rules in order to elucidate the general pattern, including\footnote{These computations require the explicit forms of the vanishing singular vectors of the staggered modules $\LogMod{1,s}$.}
\begin{align}
\IndMod{1,2} \fuse \LogMod{1,4} &= 2 \: \IndMod{1,3} \oplus \LogMod{1,5}, & \IndMod{1,2} \fuse \LogMod{1,5} &= \LogMod{1,4} \oplus \IndMod{1,6}, \notag \\
\IndMod{1,3} \fuse \LogMod{1,4} &= 2 \: \LogMod{1,4} \oplus \IndMod{1,6}, & \IndMod{1,3} \fuse \LogMod{1,5} &= 2 \: \IndMod{1,3} \oplus \LogMod{1,7}, \\
\LogMod{1,4} \fuse \LogMod{1,4} &= 4 \: \IndMod{1,3} \oplus 2 \: \LogMod{1,5} \oplus \LogMod{1,7}, & \LogMod{1,5} \fuse \LogMod{1,5} &= \IndMod{1,3} \oplus 2 \: \LogMod{1,5} \oplus \LogMod{1,7} \oplus \IndMod{1,9}. \notag
\end{align}
The module $\LogMod{1,7}$ appearing here is defined by the exact sequence $0 \rightarrow \IndMod{1,5} \rightarrow \LogMod{1,7} \rightarrow \IndMod{1,7} \rightarrow 0$, and the conditions
\begin{equation}
L_0 \ket{\phi_{1,7}} = 5 \ket{\phi_{1,7}} + \ket{\xi} \qquad \text{and} \qquad A_3 \ket{\phi_{1,7}} = \frac{-35}{3} \ket{\phi_{1,5}}, \label{eqnA3.17}
\end{equation}
where $\ket{\phi_{1,7}}$ is the logarithmic partner of $\ket{\xi} = \brac{L_{-3} - L_{-2} L_{-1} + \tfrac{1}{6} L_{-1}^3} \ket{\phi_{1,5}}$ (the non-vanishing singular vector of $\IndMod{1,5}$), and $A_3 = L_3 - L_1 L_2 + \tfrac{1}{6} L_1^3$.

The general pattern observed is best expressed as follows:
\begin{enumerate}
\item Replace each staggered module $\LogMod{1,3m+n}$ ($n=1,2$) by the direct sum $\IndMod{1,3m-n} \oplus \IndMod{1,3m+n}$ to which it is isomorphic as a vector space (but not as a Virasoro module).
\item Compute the fusion using distributivity and the na\"{\i}ve fusion rules of \secref{secConstructions1}:
\begin{equation}
\IndMod{1,s} \fuse \IndMod{1,t} = \IndMod{1,\abs{s-t}+1} \oplus \IndMod{1,\abs{s-t}+3} \oplus \ldots \oplus \IndMod{1,s+t-3} \oplus \IndMod{1,s+t-1}.
\end{equation}
\item Replace direct sums of the form $\IndMod{1,3m-n} \oplus \IndMod{1,3m+n}$ ($n=1,2$) by the corresponding staggered module $\LogMod{1,3m+n}$.  It is not hard to check that there will always be a unique way of doing this.
\end{enumerate}
It should be clear that closure under fusion requires that the spectrum of the logarithmic \cft{} we have constructed contains the modules
\begin{equation} \label{spectrum}
\IndMod{1,1}, \quad \IndMod{1,2}, \quad \IndMod{1,3k} = \IrrMod{1,3k}, \quad \LogMod{1,3k+1} \quad \text{and} \quad \LogMod{1,3k+2} \qquad \text{($k \in \ZZ_+$)}.
\end{equation}
Here, $\LogMod{1,3k+1} \cong \IndMod{1,3k-1} \oplus \IndMod{1,3k+1}$ and $\LogMod{1,3k+2} \cong \IndMod{1,3k-2} \oplus \IndMod{1,3k+2}$ as vector spaces.

We illustrate this procedure with an example.  To compute $\LogMod{1,4} \fuse \LogMod{1,5}$, first note that
\begin{equation}
\brac{\IndMod{1,2} \oplus \IndMod{1,4}} \fuse \brac{\IndMod{1,1} \oplus \IndMod{1,5}} = 2 \: \IndMod{1,2} \oplus 3 \: \IndMod{1,4} \oplus 2 \: \IndMod{1,6} \oplus \IndMod{1,8}.
\end{equation}
We infer from this the fusion rule
\begin{equation}
\LogMod{1,4} \fuse \LogMod{1,5} = 2 \: \LogMod{1,4} \oplus 2 \: \IndMod{1,6} \oplus \LogMod{1,8},
\end{equation}
where $\LogMod{1,8}$ is a staggered module with exact sequence $0 \rightarrow \IndMod{1,4} \rightarrow \LogMod{1,8} \rightarrow \IndMod{1,8} \rightarrow 0$.  We have of course checked this result through direct computation.

\section{Discussion} \label{secDiscussion}

The identification of critical percolation with a logarithmic \cft{} has received much attention recently.  Indeed, this was even argued by Cardy himself \cite{CarLog99} for a general class of disordered quenched systems with trivial partition function (that includes percolation), but without a detailed supporting \cft{} construction.  We will now compare our theory with the other logarithmic theories that have been proposed in the literature.

We first compare with the proposed theory of Read and Saleur \cite{ReaAss07} who studied a $c=0$ theory defined by the continuous limit of a $\func{\mathcal{U}_q}{\alg{sl}_2}$ XXZ spin-$\frac{1}{2}$ chain of even length at $q=e^{\ii \pi /3}$.  By analysing the associated Temperley-Lieb algebra, they deduced the existence of modules which may be identified with our $\IndMod{1,1}$, $\IndMod{1,6k-3} = \IrrMod{1,6k-3}$, $\LogMod{1,6k-1}$ and $\LogMod{1,6k+1}$, for $k \in \ZZ_+$.  These are the modules in (\ref{spectrum}) with \emph{odd} second subscript label.  This forms a fusion subring of that which we have computed in that the fusion rules given in \cite{ReaAss07} agree with the appropriate restriction of ours (and close).  It is worth mentioning however that their proposed theory does not contain a field that may be identified with $\phi_{1,2}$, and so cannot explain the crossing probability computation of Cardy.

This contrasts with the fusion ring proposed by Pearce and Rasmussen \cite{RasFus07}.  This was deduced from numerical studies of an integrable lattice model of critical percolation, defined in their prior work with Zuber using Temperley-Lieb algebras to obtain lattice constructions of logarithmic extensions of minimal models \cite{PeaLog06}.  Tellingly, they propose fusion rules for modules corresponding to \emph{all} fields in the extended Kac table ($\phi_{r,s}$ with $r,s \in \ZZ_+$).  This is necessitated by their assumption that both $\IndMod{1,2}$ and $\IndMod{2,1} = \IrrMod{2,1}$ are present in the theory.  Cardy's crossing probability result only requires the former to be present, so our fusion ring may be identified with a subring of theirs, in fact, the subring which they refer to as the ``vertical fusion algebra''.

This ``extended'' fusion ring, as reported by Pearce and Rasmussen, is in turn identical to a subring of the ring previously proposed by Eberle and Flohr \cite{EbeVir06}, based on extensive computations using the algorithm of Nahm-Gaberdiel-Kausch (as ours are).  It is clear, however, that their starting assumption is that of irreducibility (which we rejected in \secref{secIntro}).  They assume that every irreducible module in the extended Kac table is present, though the trivial irreducible module $\IrrMod{1,1} = \IrrMod{1,2}$ is noted to decouple from the fusion ring obtained and is removed.  The indecomposable modules $\IndMod{1,1}$ and $\IndMod{1,2}$ are then added to the theory, as they are found to occur as submodules of indecomposable modules generated by fusing the above irreducibles (although they do not seem to add $\IndMod{1,4}, \IndMod{1,5}, \ldots$, which also appear as such submodules).  The spectrum of their theory is therefore even richer than that of Pearce and Rasmussen, and we obviously again find agreement between our fusion ring and theirs (when restricted to our spectrum).

Eberle and Flohr were also able to further characterise the modules appearing in their theory by determining certain parameters $\beta$ (originally discussed in \cite{GabInd96}) associated to the staggered modules (which we denote by $\LogMod{r,s}$).  In particular, they give $\beta = \tfrac{-1}{2}$ and $\beta = \tfrac{-5}{8}$ for $\LogMod{1,4}$ and $\LogMod{1,5}$ respectively, agreeing with our \eqnDref{eqnL1.14}{eqnL2.15}, respectively.  However, the parameters (sometimes they give two) that they determine for more general $\LogMod{r,s}$ are \emph{not} invariants of the module itself, so they are difficult to independently verify.  We have computed the invariant parameter $\beta = -\tfrac{35}{3}$ (and there is only ever one \cite[Thm.\ 5.12]{RohRed96}) for $\LogMod{1,7}$ in \eqnref{eqnA3.17}, and verified that this value is always found, regardless of which fusion rule is used to generate this module (this we checked with $\LogMod{1,4} \fuse \LogMod{1,4}$, $\LogMod{1,5} \fuse \LogMod{1,5}$ and $\IndMod{1,2} \fuse \IndMod{1,6}$).

To elaborate, the first part of \eqnref{eqnA3.17} only defines $\ket{\phi_{1,7}}$ modulo the kernel $\mathcal{K} \subset \LogMod{1,7}$ of $L_0 - 5 \id$.  This kernel is the three-dimensional subspace of grade-$3$ descendants of the \hws{} $\ket{\phi_{1,5}}$ of $\LogMod{1,7}$ (not to be confused with the non-\hws{} $\ket{\phi_{1,5}}$ in $\LogMod{1,5}$).  Eberle and Flohr define their parameter $\beta$ to be the multiple of $\ket{\phi_{1,5}}$ obtained by letting $L_1^3$ act on $\ket{\phi_{1,7}}$.  However, $L_1^3$ does not act trivially on $\mathcal{K}$, so their $\beta$ depends upon which particular $\ket{\phi_{1,7}}$ they have chosen (and so can take a continuous range of values).  To get an invariant $\beta$, one must act on $\ket{\phi_{1,7}}$ with a (raising, grade $3$) operator which annihilates $\mathcal{K}$.  There is of course only one such operator (up to scalar multiples) and it is found by taking the singular vector $\ket{\xi} \in \mathcal{K}$ (whose logarithmic partner is $\ket{\phi_{1,7}}$) and removing $\bra{\phi_{1,5}}$ from $\bra{\xi}$.  This is the operator we have denoted by $A_3$ in \eqnref{eqnA3.17}, obtaining $\beta = \tfrac{-35}{3}$.

In fact, the above analysis makes it clear that this invariant is (for a general staggered module $\LogMod{r,s}$) nothing but
\begin{equation}
\beta_{r,s} = \braket{\chi}{\phi_{r,s}},
\end{equation}
where $\ket{\chi}$ is the non-vanishing singular vector in the highest weight (indecomposable) submodule of $\LogMod{r,s}$ and $\ket{\phi_{r,s}}$ is its logarithmic partner state (which is not in this submodule).  (We can now identify the constant $C$ of \eqnref{eqnLogCoup14} with $\beta_{1,4}$---see also \eqnref{eqnL1.14}).  $\beta_{r,s}$ therefore quantifies the degree to which the highest weight submodule is coupled to its logarithmic partner module.  We therefore call this staggered module invariant the \emph{logarithmic coupling}.  It is now evident that this invariant in fact scales with the (square of the) normalisation of the singular vector $\ket{\chi}$.  We always normalise $\ket{\chi}$ (and hence $\beta_{r,s}$) so that the term with the single (most negative) Virasoro mode has coefficient $1$ (and we order the modes in the other terms by non-decreasing index).

To summarise the comparisons made thus far, we have identified the fusion ring of Read and Saleur as a subring of ours which does not contain $\phi_{1,2}$, and so their theory does not provide a formalism in which to understand Cardy's crossing probability result.  On the other hand, the fusion rings proposed by Pearce--Rasmussen and Eberle--Flohr contain our fusion ring as a subring, and so are sufficiently rich to explain Cardy's result.  Unfortunately, the spectral excess (over what is strictly necessary for the non-triviality of the $\phi_{1,2}$ four-point function) in these enlarged fusion rings clashes with conformal invariance.  This is due to a subtlety involving logarithmic couplings, and follows from an argument originally due to Gurarie and Ludwig \cite[App.\ A]{GurCon04} which we briefly outline.

As we have shown, if the theory contains the module $\IndMod{1,2}$ (or $\IndMod{1,3} = \IrrMod{1,3}$), then fusion generates the module $\LogMod{1,5}$.  If $\IndMod{2,1} = \IrrMod{2,1}$ is also present, then we additionally generate a module $\LogMod{3,1}$ with exact sequence $0 \rightarrow \IndMod{1,1} \rightarrow \LogMod{3,1} \rightarrow \IndMod{3,1} \rightarrow 0$.  $\ket{\phi_{3,1}}$ has conformal dimension $2$ (\tabref{tabExtKacc=0}), and satisfies (compare with \eqnref{eqnL2.15})
\begin{equation}
L_0 \ket{\phi_{3,1}} = 2 \ket{\phi_{3,1}} + L_{-2} \ket{0} \qquad \text{and} \qquad L_2 \ket{\phi_{3,1}} = \frac{5}{6} \ket{0}.
\end{equation}
If, however, the conformal invariance of the vacuum is used to compute the correlation function $\corrfn{\func{\phi_{1,5}}{z} \func{\phi_{3,1}}{w}}$, one finds that global invariance under $L_{-1}$ and $L_0$ fix the form of this function completely (as with \eqnref{eqnCF15.15}), but this form does \emph{not} satisfy the $L_1$-invariance constraint, essentially because the logarithmic couplings $\beta_{1,5} = \tfrac{-5}{8}$ and $\beta_{3,1} = \tfrac{5}{6}$ are not equal (whilst the respective dimensions of the generating states of these modules are).  The conclusion is then that one cannot have both $\IndMod{1,2}$ and $\IndMod{2,1}$ in the theory simultaneously\footnote{We stress that this argument proves that one cannot augment the theory introduced above by the \emph{module} $\IndMod{2,1}$, whose \hws{} has dimension $\tfrac{5}{8}$ (\tabref{tabExtKacc=0}).  It does not rule out the possibility of consistently adding a primary field of this dimension.  However, the \hws{} corresponding to such a field cannot have a vanishing singular vector at grade $2$.  We expect that an extended algebra approach will be able to determine whether such augmentations are also forbidden.}.

The work of Gurarie and Ludwig (detailed in \cite{GurCon04}) is of considerable relevance to our construction.  Their view is to construct $c=0$ logarithmic theories by assuming that the theory satisfies a particular set of carefully chosen \opes{}.  Many of these involve a logarithmic partner field to $T$ which we can identify in our theory with $\phi_{1,5}$.  The focus of their work was not to construct the theory from its fusion ring (fusion processes are not treated there), but to investigate the consequences of extending the Virasoro algebra by the modes of this partner field.  It is very interesting to see that their partial extension already allows them to determine ``anomaly numbers'' $b$ \cite{GurCTh99}, which coincide with the logarithmic couplings $\beta$ we have discussed above in the two cases they treat, $\tfrac{-5}{8}$ and $\tfrac{5}{6}$.

As noted above, the staggered modules $\LogMod{1,5}$ and $\LogMod{3,1}$ cannot both be simultaneously present in a consistent \cft{}.  Gurarie and Ludwig realised that this means that there are (at least) two distinct logarithmic theories that one can construct at $c=0$ (and we venture that Read and Saleur's XXZ spin chain theory \cite{ReaAss07} perhaps leads to a third).  Moreover, they identified in \cite{GurCon02} the one containing $\LogMod{1,5}$ ($\beta_{1,5} = \tfrac{-5}{8}$) as realising polymers and that containing $\LogMod{3,1}$ ($\beta_{3,1} = \tfrac{5}{6}$) as realising percolation (however, this identification was not reaffirmed or refuted in the sequel \cite{GurCon04}).  Contrarily, we maintain that percolation must involve $\LogMod{1,5}$ (and more fundamentally, $\IndMod{1,2}$), hence percolation has $\beta_{1,5} = \tfrac{-5}{8}$.

In finishing, let us reemphasise the essential aspects of our construction.  To explain Cardy's result, we deform the $\MinMod{2}{3}$ model by breaking the Kac symmetry $\ket{\phi_{1,1}} = \ket{\phi_{1,2}}$.  The simplest way of doing this is by rendering the two modules reducible (but indecomposable), each differently, by allowing one of the primitive singular vectors in each module to be physical (non-vanishing).  The proper choices, $L_{-2} \ket{\phi_{1,1}} \neq 0$ and $L_{-1} \ket{\phi_{1,2}} \neq 0$, which transform $\IrrMod{1,s}$ into $\IndMod{1,s}$ ($s=1,2$), are fixed by the physics.  (Note that this starting point fits naturally with the point of view that percolation is to be regarded as a limiting theory with $c \rightarrow 0$:  In this picture, the natural modules to consider are precisely our $\IndMod{1,s}$.)  In a second step, we have shown how the logarithms arise naturally---without further input---from these assumptions.

Our formalism is also well-suited to interpreting and consolidating the results of Gurarie and Ludwig.  In particular, we hope to use the framework we have developed to investigate the (partial) extended algebra approach that they have pioneered.  Furthermore, it is clear that our constructions may be easily adapted to defining logarithmic theories corresponding to every minimal model.  We expect that these theories will prove to be the correct framework in which to explain the occurrence of non-local observables corresponding to fields outside the (standard) Kac table in other critical models (the Ising model \cite{ArgNon02} for example).

\section*{Acknowledgements}

We thank Yvan Saint-Aubin for discussions and critical comments on the manuscript, and Matthias Gaberdiel for clarifying certain aspects of the Nahm-Gaberdiel-Kausch algorithm for fusion.

\end{document}